\documentstyle[11pt,aaspp4]{article}

\tighten

\begin{document}

\title{CURVATURE OF THE UNIVERSE AND OBSERVED GRAVITATIONAL
       LENS IMAGE SEPARATIONS VERSUS REDSHIFT}

\author{Myeong-Gu Park\altaffilmark{1}}
\affil{Department of Astronomy and Atmospheric Sciences,
       Kyungpook National University, Taegu 702-701, KOREA;
       mgp@bh.kyungpook.ac.kr}
\author{J. Richard Gott III}
\affil{Department of Astrophysical Sciences, Princeton University,
       Princeton, NJ 08544; jrg@astro.princeton.edu}

\altaffiltext{1}{also at Department of Astrophysical Sciences, 
                 Princeton University}

\begin{abstract}
In a flat, $k=0$ cosmology with galaxies that approximate
singular isothermal spheres, gravitational lens image separations
should be uncorrelated with source redshift. But in an open $k=-1$
cosmology such gravitational lens image separations become
smaller with increasing source redshift. The observed separations
do become smaller with increasing source redshift but the effect
is even stronger than that expected in an $\Omega=0$ cosmology.
The observations are thus not compatible with
the ``standard'' gravitational lensing statistics model in a flat
universe. We try various open and flat cosmologies, 
galaxy mass profiles, galaxy merging
and evolution models, and lensing aided by clusters to 
explain the correlation. 
We find the data is not compatible with any of these
possibilities within the 95\% confidence limit,
leaving us with a puzzle. 
If we regard the observed result as a statistical fluke,
it is worth noting that we are about twice as likely to
observe it in an open universe (with $0<\Omega<0.4$) as
we are to observe it in a flat one.
Finally, the existence of an observed multiple image lens 
system with a source at $z=4.5$
places a lower limit on the deceleration parameter: 
$q_0 > -2.0$.

\end{abstract}


\section{Introduction}

The list of multiple image gravitational lens systems
has been growing steadily since the discovery of the first
lens system (Walsh, Carswell, \& Weymann 1979). 
At present, about thirty multiple image systems
are confirmed or very likely to be gravitationally lensed systems
(see e.g., Surdej \& Soucail 1994; Keeton \& Kochanek 1996).
These lens systems can provide us with information
about the universe as a whole and the mass distribution 
within. 

Turner, Ostriker, \& Gott (1984; hereafter TOG) did extensive
studies on the statistical nature of gravitational lenses and their
implications for cosmology and galaxy formation. One of the results of
this work was that the mean image separations of lens systems
have different dependences
on the source redshift in different cosmologies and that it may
therefore be possible to measure the curvature of the universe
directly. Gott, Park, \& Lee
(1989; hereafter GPL) explored the lens statistics in 
more general cosmologies where the cosmological constant
$\Lambda$ is not zero. They showed that the then-available data
ruled out extreme closed models having antipodal
redshift $z_{antipode} < 3.5$, and the deceleration
parameter $q_0 > -2.3$.

As the list of lenses grows, it has been applied to variety of problems.
One prominent application is to place limits on the cosmological constant. 
With the observed galaxy mass
distribution and number density, a universe with large cosmological
constant should produce more multiple image systems than are
actually observed. This has placed steadily improving 
limits on the cosmological constant:
$\Omega_\Lambda \lesssim 0.95$ (Fukugita et al. 1992) or
$\Omega_\Lambda \lesssim 0.66$ (Kochanek 1996) where
$\Omega_\Lambda \equiv \Lambda/3H_0^2$ and $H_0$ the Hubble constant.
This limit is already strong enough to place telling constraints
on an otherwise appealing cosmological model ($\Omega+
\Omega_\Lambda =1$, $k=0$, see Ostriker \& Steinhardt 
1995 for summary) where $\Omega = 8\pi\rho_0/3H_0^2$.
In addition, Maoz and Rix (1993) investigated 
the effects of the mass distribution in
E/S0 galaxies and concluded that the HST snapshot survey data
requires E/S0 galaxies to have significant halos.  
Further studies of galaxy merger/evolution show that 
only some specific
merger models can be rejected, and the above limit on $\Lambda$ is not
affected (Rix et al. 1994; Mao \& Kochanek 1994). 
However, most applications of gravitational lensing statistics 
do assume specific mass (or
velocity dispersion) distributions for lensing galaxies,
e.g., a Schechter luminosity function and a luminosity-velocity
relation, and specific number density distributions, e.g., a constant
comoving density of galaxies.   

In this work, we focus on the image separations versus the source
redshift of the current multiple image lens systems to see whether
it is consistent with the `standard' lensing statistics models. We find
that the image separations are strongly negatively correlated with
the source redshift, which is incompatible with the `standard'
lensing statistics model in a flat universe. We explore possible
causes to see if this correlation can be explained. We also update the
limit on the deceleration parameter $q_0$ with the current data.

\section{Observed Multiple Image Lens Systems}

The list of multiple image QSO and radio sources has
grown through systematic optical and radio surveys and
through serendipitous discoveries. Keeton and Kochanek
(1996; also see Surdej \& Soucail 1994) 
summarize the data on the 29 relatively secure multiple
image lens systems. 
They classify these systems into three grades of secureness:
class A for ``I'd bet my life this is a lens.'',
class B for ``I'd bet your life this is a lens.'',
and class C for ``You should worry if I'm betting your life.'',
all of which (A, B, and C) show convincing spectral similarities
and identical redshifts (Table 1 for references) 
and have separations which are either quite
similar to 0957 which is surely a lens (having the time delay
between its two images measured recently [Kundi\'c et al. 1996])
or smaller separations.
Also, note that the largest separation lens system 2345
(in class C) now has more observational support 
for being a true lens---its lens has been found (Fischer et al. 1994).
In this work, we use all 20 systems in this list (A, B, and C)
which have a known source redshift. 
The system 2237+0305 (Huchra et al. 1985)
is not included because in that system the source (QSO)
was found after the lens. Such systems would have different statistical
properties than systems where the source is discovered first.
These 20 systems are listed in Table 1 and 
their maximum image separations, $\Delta\theta$, are 
plotted against source redshift, $z_s$, in Figure 1
(class A as circles, class B as triangles, and class C as
crosses). They show visually quite a strong negative correlation
between $\Delta\theta$ and $z_s$. The major source of this
correlation is a number of small redshift ($z_s\lesssim2$), large separation 
($\Delta\gtrsim4\arcsec$) lens systems (2345, 1120, 0240,
0957, 1429) and large redshift 
($z_s\gtrsim3.5$), small separation ($\Delta\lesssim1\arcsec$),
lens systems (1208, J03.13). (This effect was noted by Gott [1997]. 
The original data [7 QSO's] in GPL showed no significant correlation.)

Of course there is always a possibility of contamination by ``false''
lenses, i.e., observing real physical pairs of QSO's, 
at wide separation at small source redshift due to
quasar clustering (which might be larger at low source redshift).
We can roughly estimate how many QSO physical pairs
might be expected to show up as ``false'' lenses. 
Djorgovski (1991) lists
3 quasar pairs (or triplets) with arcmin-scale separations and 
$\Delta V_{rest} < 1000 \hbox{km}\:\hbox{s}^{-1}$ where 
$\Delta V_{rest}$ is the redshift difference between quasars:
QQ 0107-025 AB ($z=0.954$ and $\Delta\theta=77\arcsec$),
QQ 1146+111 BC ($z=1.012$ and $\Delta\theta=157\arcsec$),
and Hoag 1,2,3 ($z=2.049$ and $\Delta\theta=121\arcsec$, 
$128\arcsec$, and $214\arcsec$).
This number roughly agrees with the covariance function 
$w(\theta) \propto \theta^{-0.8}$ expected for gravitational clustering
with the average comoving density of quasars 
$<\rho> \simeq 1000\:{\hbox{Gpc}}^{-3}$ and a correlation length
$r_0 \simeq 10h^{-1}\:\hbox{Mpc}$. 
From the power law shape of $w(\theta)$, the existence of
two QSO pairs within $128\arcsec<\Delta\theta<256\arcsec$
implies that we would expect to see roughly 0.06 QSO pairs in the interval
$0\arcsec<\Delta\theta<8\arcsec$. Hence, 
the contamination would be unimportant if QSO pairs follow the
covariance function expected for the hierarchical clustering. 
However, Djorgovski (1991) also lists
3 QSO pairs with arcsec-scale separation and 
$\Delta V_{rest} < 1000 \hbox{km}\:\hbox{s}^{-1}$, which are about
two orders of magnitude over-abundant relative to the prediction of 
hierarchical clustering: 
PKS 1145-071 AB ($z=1.345$ and $\Delta\theta=4\farcs2$), 
0151+048 AB ($z=1.91$ and $\Delta\theta=3\farcs3$), and 
QQ 1343+266 AB ($z=2.030$ and $\Delta\theta=9\farcs5$).
Among these, only the 1145 AB pair has the required 
spectral similarities in the optical to be confused with a lens
(Djorgovski et al. 1987) and which is within the
$4\arcsec<\Delta\theta<8\arcsec$ interval.
If we regard 0957 and 2345 as the only
proven gravitational lensed cases with $\Delta\theta>4\arcsec$, 
the a priori probability that a given 
QSO pair with $4\arcsec<\Delta\theta<8\arcsec$
with similar spectra is a gravitational lens rather than
a physical pair is 2 in 3 (because of the decided cases
0957 and 2345 are lenses while 1145 is not).
Hence, the probability that 
all three remaining systems (1120, 0240, 1429) in Table 1 
within $4\arcsec<\Delta\theta<8\arcsec$
are just physical pairs (even if one disregards all other observations)
is $1/3^3$, which is less than 5\%. The probability that
two specific systems (for example, 1120 and 0240) are physical pairs
is $1/3^2$, and the probability that one specific system
is physical pair is $1/3$. 
The quantitative aspects of possible
contamination of a few ``false'' cases is discussed later.
So for the time being, we are treating all 20 cases in Table 1
as real gravitational lens systems.

\section{Geometry of the Universe}

\subsection{Flat and Open Cosmological Models}

One of the most important cosmological parameters is the curvature
of the universe. The Friedmann Big Bang models admit three
solutions: universes that are (1) flat, $k=0$, with a
Euclidean three sphere space geometry $R^3$ at fixed epoch,
(2) closed, $k=+1$, with a $S^3$ three space geometry at
fixed epoch, and (3) open, $k=-1$, with a hyperbolic $H^3$
three space geometry at fixed epoch (Misner, Thorne, \& Wheeler
1973). We would very much like to know whether our universe
is flat, closed or open so direct measurement of the curvature is
extremely important. Models with $\Omega+\Omega_\Lambda<1$
are open ($k=-1$), models with $\Omega+\Omega_\Lambda=1$ are
flat ($k=0$), and models with $\Omega+\Omega_\Lambda>1$ are
closed ($k=+1$).

Flat, $k=0$ models (with $\Omega=0.4$, $\Omega_\Lambda=0.6$) are
popular with many people (cf. Ostriker \& Steinhardt 1995)
because it could be produced naturally in any inflationary
scenario where significantly more than 67-e-folds of inflation
occurs and (other than the theoretical problems with a finite
$\Lambda$ term) would require no fine tuning of parameters.
But there are also open ($k=-1$) inflationary models. Open
inflationary universes, as suggested by Gott (1982), are 
created naturally during the decay of an initial metastable
inflationary state. Individual bubble universes are created 
which have an open geometry, with a negative curvature
inherited from the bubble formation event. Inflation continues
within the bubble for approximately 67 e-folding times,
creating a universe with a radius of curvature $\exp(67)$ times
larger than the wavelength of the microwave background photons,
and which is uniform except for quantum fluctuations
(cf. Gott \& Statler 1984; Gott 1986; Gott 1997). The single-bubble
open inflationary model (Gott 1982) has come under increased
discussion recently because of a number of important developments.
On the theoretical side, Ratra \& Peebles (1994, 1995) have shown
how to calculate the random quantum fluctuations in the $H^3$
hyperbolic geometry. This is very important since it allows 
predictions of fluctuations in the microwave background.
Bucher, Goldhaber, \& Turok (1995a,b) have done similar
calculations, as well as Yamamoto, Sasaki, \& Tanaka (1995).
Importantly, they have explained that the fine tuning in these
models is only ``logarithmic'' and, therefore, not so serious.
Linde (1995) has shown how there are reasonable potentials which
could produce such open universes, indeed different open
bubble universes with different values of $\Omega$.

The inflationary power spectrum with CDM (Bardeen, Steinhardt, \&
Turner 1983) has been amazingly successful in explaining 
the qualitative features of observed galaxy clustering
including great walls and great attractors
(cf. Geller \& Huchra 1989; Park 1990a,b; Park \& Gott 1991).
The amount of large scale power seen in the observations
suggests an inflationary CDM power spectrum with 
$0.2 < \Omega h < 0.3$ (Maddox et al. 1990; Saunders et al. 1991;
Park et al. 1992; Shectman et al. 1995; Vogeley et al. 1994).
A number of recent estimates of $h$ 
have been $> 0.55$ (i.e., $h=0.65\pm0.06$ [Riess, Press, \& Kirshner
1995), $0.68 \leq h \leq 0.77$ [Mould \& Freedman et al. 1996],
$0.55 \leq h \leq 0.61$ [Sandage et al. 1996], and
$h = 0.67 \pm 0.06$ from the time delay of 418 days observed in 0957
[Kundi\'c et al. 1996] using the best model by 
Grogin \& Narayan [1996]). Ages of globular cluster stars
have a 2$\sigma$ lower limit of about 11.6 billion years
(Bolte \& Hogan 1995); if the age of the universe $t_0 \geq 11.6$
billion years, we require $h < 0.56$ if $\Omega=1$ and
$\Omega_\Lambda=0$, but a more
acceptable $h < 0.65$ if $\Omega=0.4$, $\Omega_\Lambda=0$.
Models with low $\Omega$ but $\Omega+\Omega_\Lambda=1$ are also
acceptable. With the COBE normalization there is also the problem
that with $\Omega=1$ and $\Omega_\Lambda=0$, 
$(\delta M/M)_{8 \:{\hbox{h}}^{-1}
\:\hbox{Mpc}}= 1.1-1.5$ and this would require galaxies to be
anti-biased [since for galaxies $(\delta M/M)_{8 \:{\hbox{h}}^{-1}
\:\hbox{Mpc}}= 1$] and this would also 
lead to an excess of large-separation
gravitational lenses over those observed (Cen et al. 1994).
These things have forced even enthusiasts of $k=0$ models to move to
models with $\Omega < 1$ but with a cosmological constant
so that $\Omega + \Omega_\Lambda = 1$ and $k=0$.

\subsection{Gravitational Lensing Curvature Test}

In this paper, we will discuss a curvature test based on 
gravitational lens image separations as a function of source redshift.
Studies on statistics of lensing show that if a source
is lensed by a singular isothermal sphere (SIS) galaxies, randomly
distributed in the universe with constant comoving density,
the mean separation---averaged over all possible lenses
at different distances---of multiple images in a flat universe should be
constant independent of the source redshift (solid line in Fig. 2a). 
However, the mean separation will decrease with source 
redshift in an open universe (dotted and dashed lines in Fig. 2a) 
and increase in a closed universe (TOG, GPL). In an
open universe the volume increases faster with redshift
than in a flat universe,
and the source is more likely to be lensed by lensing galaxies at
larger distances, which produces smaller image separations, 
and vice versa for a closed universe.
This applies to lensing by galaxies and/or clusters both of which
are well approximated by SIS. Also this test is independent
of individual values of $\Omega$ and $\Omega_\Lambda$ when
the universe is flat ($\Omega+\Omega_\Lambda=1$). 

This is quite important because it is a pure curvature test that
distinguishes a $k=0$ cosmology from a $k=-1$ cosmology.
We have tests for $\Omega$: i.e., peculiar velocities are
proportional to $\Omega^{0.6}/b$ where $b$ is the bias parameter,
and power on large scales in galaxy scale clustering measures
$\Omega h$ where $h= H_0 / 100~\hbox{km}~\hbox{s}^{-1}~\hbox{Mpc}^{-1}$.
But these tests do not distinguish between a model with
$\Omega<1$, $\Omega_\Lambda=0$ which is open ($k=-1$) and
a model with the same value of $\Omega$ but with 
$\Omega+\Omega_\Lambda=1$ which is flat ($k=0$). 

How can we distinguish between the $\Omega = 0.3-0.4$,
$\Omega_\Lambda = 0$, $k=-1$ models, and the $\Omega=0.3-0.4$,
$\Omega_\Lambda = 0.6-0.7$, $k=0$ models? They produce galaxy
clustering and masses of groups and clusters that are virtually
indistinguishable. Turner (1990) and Fukugita, Futamase, \&
Kasai (1990) showed that a flat $\Omega_\Lambda =1$ model
produces about 10 times as many gravitational lenses than a 
flat model with $\Omega = 1$. By comparing with the observed
number of lenses, Kochanek (1996) was able to set a 95\%
confidence lower limit of $0.34<\Omega$ in flat models where
$\Omega+\Omega_\Lambda=1$, and a 90\% confidence lower limit
$0.15<\Omega$ in open models with $\Omega_\Lambda=0$. Thus,
extreme-$\Lambda$ dominated models are ruled out by producing too
many gravitational lenses. Another possibility is future
data on the cosmic background radiation for spherical harmonic
modes from $l=2$ to $l=500$: the $\Omega=1$, $\Omega_\Lambda=0$
model reaches its peak value at $l\simeq200$; an $\Omega=0.3$,
$\Omega_\Lambda=0.7$ model reaches its peak value at $l\simeq200$
(Ratra et al. 1995); while an $\Omega=0.4$, $\Omega_\Lambda=0$
model reaches its peak value at $l\simeq350$ (Ratra \& Sugiyama 1995).
This can be measured by the MAP and COBRAS/SAMBA satellites which
will measure this range with high accuracy.
The test in this paper (gravitational lens separations as a
function of source redshift) is also able in principle to
differentiate between an $\Omega=0.4$, $\Omega_\Lambda=0$,
$k=-1$ model and an $\Omega=0.4$, $\Omega_\Lambda=0.6$,
$k=0$ model. 

\subsection{Curvature Test Results}

For our gravitational lensing curvature test
we first estimate the probability of producing the observed
correlation by chance in a flat 
universe where the distribution of the separations is
expected to be independent of the source redshift. 
Since we don't assume 
any specific distribution of image separations at a given redshift, 
we use Spearman's rank correlation test. The Student-t distribution 
gives the approximate probability for the random distribution 
to have stronger than a given correlation (Press et al. 1992).  
Whenever there are ties, midranks are used. 
We checked this probability against Monte-Carlo simulations
and they agree well. The two-sided probability 
(of observing either a positive correlation or negative 
correlation as strong as that observed in Fig. 1) 
in a flat universe with SIS galaxies
is $P = 0.012$ (Table 2). This confirms
the visual impression that the distribution is significantly
(negatively) correlated with source redshift. 
We also divide the sample into three redshift intervals,
$[0,2]$, $[2,3]$, and $[3,\infty]$ and apply the Kolmogorov-Smirnov test.
The distribution of separations
in $[0,2]$ and $[2,3]$ are not significantly different.
However, those in $[0,2]$ and $[3,\infty]$ are statistically
different with 95\% confidence. 
If we are in a flat universe, this is a very special sample.

To see the possible effect of any ``false'' cases,
we repeat the Spearman test for the data set where some cases
are excluded intentionally. For example, if we exclude any two
large separation ($4\arcsec<\Delta\theta<8\arcsec$) systems (except 0957,
of course), the probability is small $P\leq0.051$.
Excluding three systems, for example, 1120, 0240, 1429, increases
the probability only to $P=0.063$.
Similarly even if the most favorable 
large separation and small separation case
are excluded (1120 and 1208), the probability is still small $P=0.029$. 
Only when two large separation and one small separation cases (1120, 0240,
and 1208) are excluded, is $P=0.066$. 
On the other hand, if two of the largest
reshift cases (0952 and 1208, both class B) are excluded, 
the probabilty becomes quite significant, $P=0.098$. 
So we conclude that
three or more largest separation ``false'' cases 
or two or more largest redshift ``false'' cases are needed to
change the incompatibility
of the observed data with the standard lensing statistics model
in a flat universe at the $\sim 95\%$ confidence level. 

If it is not just a statistical fluke, what could
be responsible for this correlation? We first check
if negative curvature can create this strong a trend. We try two
open universes: an $\Omega=0$, $\Omega_\Lambda=0$ empty universe and
an $\Omega = 0.4$, $\Omega_\Lambda = 0$ open universe. The
mean image separation, $<\Delta\theta>$, is calculated 
as a function of the source redshift $z_s$ 
(in Fig. 2a: a dashed line for $\Omega=0$, $\Omega_\Lambda=0$ and
a dotted line for $\Omega=0.4$, $\Omega_\Lambda=0$).
We then divide the observed image
separations $\Delta\theta_i^{obs}$ by 
expected mean separation $<\Delta\theta>(z_s)$.
If the correlation is due to the curvature, these ``normalized''
separations should not show any correlation with $z_s$. 
However, Spearman tests
indicate that in both the empty and $\Omega = 0.4$ open universes,
significant correlations still exist, and the probability 
that the data could be randomly drawn from these empty and 
$\Omega=0.4$ models is $P=0.030$ and $P=0.019$, respectively (Table 2). 
So although negative curvature lessens the strength of the
correlation, it alone cannot fully explain the correlation.
We also test for the possible effect of ``false'' cases in $\Omega=0.4$
open universe. Exclusion of 1120 from the data set
increases the probability to $P=0.039$ and of 1120 and 1429
to $P=0.048$, while exclusion of 1120 and 0240 increases
the probability to $P=0.080$. Also, exclusion of 1208
(smallest separation) increases the probability to $P=0.059$,
just above 5\% level, although the correlation still exists. 
This is higher than the probability for
the flat universe because some of the negative correlation would
be explained by the curvature of the universe. 

It is also worth noting that
if this is just a statistical fluke, we are about twice as
likely to see it in an open universe (with $0\leq\Omega\leq0.4$)
than in a flat universe (with $\Omega+\Omega_\Lambda=1$). 

\section{Mass Profile}

Other factors that can affect the distribution of image separations
include the density profile of the lenses. The density profile of SIS
produces a constant bending angle regardless of the impact parameter, and
the distribution of image separations is independent of the source
redshift if lenses are uniformly distributed in a flat universe. 
If the density profile is steeper than SIS, image separations
decrease as the source redshift increases (TOG, GPL). 
To access the effect of a steeper density distribution, we try the
extreme case of point mass lenses. 

In a flat universe, the mean separation of images produced by point
mass lenses decreases by a factor of 0.82 from 
$z_s = 1.5$ to $z_s = 4.0$. We again calculate 
$<\Delta\theta>(z_s)$ (Fig. 2b),
normalize the observed image separation $\Delta\theta_i^{obs}$
with it, and test for any
correlation. The probability of finding either
a positive correlation or a negative correlation
as large as observed in this model is $P=0.030$ (Table 2). 
So even this most extreme density profile
cannot explain the correlation.

\section{Galaxy Merger and Infall}

The next possibility is that of evolution of the lenses
(galaxies). If the number density or the mass of the lenses changes over
cosmic time scales, this introduces a dependence of image separations
on the source redshift: If the comoving number density increases with
redshift, that is, more lenses per comoving volume at higher redshift, 
the mean separation decreases with source redshift. 
If the lens mass decreases
with redshift, the mean separation again decreases with redshift.

Following GPL notation, we represent a Robertson-Walker metric as
\begin{equation}
   ds^2 = -dt^2 + \frac{a^2(t)}{a_0^2}
          [ a_0^2 d\chi^2 + a_0^2 S^2(\chi)
                            (d\theta^2 + \sin^2\theta d\phi^2) ],
\end{equation}
where $S(\chi)=\chi$ for a flat universe, 
$S(\chi)=\sin(\chi)$ for a closed universe,
$S(\chi)=\sinh(\chi)$ for an open universe.
The comoving distance $\chi$ is related to $z$ through
\begin{equation}
   \chi = \Delta \int_0^z \left[
          \Omega(1+t)^3+(1-\Omega-\Omega_\Lambda)(1+t)^2+\Omega_\Lambda
          \right]^{-1/2} dt.
\end{equation}
Then $S(\chi)$
is equal to the proper motion distance times $\Delta$
where $\Delta = |\Omega+\Omega_\Lambda-1|^{1/2}$ 
in a closed or open universe and $\Delta=1$ in a flat universe
(see e.g., Kochanek 1993). Here, $\Omega$ and $\Omega_\Lambda$
represent the values observed at the present epoch.
The scale factor of the universe $a(t)$ has a present value
$a_0 = c H_0^{-1} \Delta^{-1}$ where $c$ is the speed of light.

The probability of lensing, 
in the general case where lenses evolve is given by 
\begin{equation}
   \tau = \pi a_0^3 n_0 \alpha_0^2 
   \int_{0}^{\chi_s} \frac{n(\chi_l)}{n_0}
   \left[ \frac{\alpha(\chi_l)}{\alpha_0} \right]^2
   \frac{S^2(\chi_s-\chi_l)S^2(\chi_l)}{S^2(\chi_s)} d\chi_l,
\end{equation}
where $n(\chi_l)$ is the comoving density,
$\alpha(\chi_l)$ the bending angle of the SIS lenses at the
distance $\chi_l$.
The subscript `0' refers to
values at present. The mean angular separation as a function of
the comoving distance of the source, $\chi_s$ ($z_s$), is 
\begin{equation}
   <\Delta\theta> = 2 \alpha_0
   \frac
   {\int_{0}^{\chi_s} \frac{n(\chi_l)}{n_0}
   \left[ \frac{\alpha(\chi_l)}{\alpha_0} \right]^3
   \frac{S^3(\chi_s-\chi_l)S^2(\chi_l)}{S^3(\chi_s)} d\chi_l} 
   {\int_{0}^{\chi_s} \frac{n(\chi_l)}{n_0}
   \left[ \frac{\alpha(\chi_l)}{\alpha_0} \right]^2
   \frac{S^2(\chi_s-\chi_l)S^2(\chi_l)}{S^2(\chi_s)} d\chi_l}.
\end{equation}

Merging between galaxies and the infall of surrounding mass onto galaxies 
are two possible processes that can change the either comoving 
density of galaxies and/or their mass. 
The effects of galaxy merging or evolution have been studied by Rix et
al. (1994) and Mao and Kochanek (1994). They focused on the lensing
probability and the limits on the cosmological constant.
Under the generic relation between the velocity dispersion
and mass of early-type galaxies, they find merging and/or evolution do
not significantly change the statistics of lensing. 

We try three merger/infall models. The first merger model
is that of Broadhurst et al. (1992) 
which was originally motivated by the
faint galaxy population counts. The exact nature of
excess of faint galaxy counts is uncertain at present.
Excess counts at large redshift may indicate that 
one is just seeing pieces, like giant HII regions (Colley et al. 1996)  
[with appropriate $K$ corrections] of
already formed galaxies rather than
galaxy mergers. In this case, the lensing
statistics would be unaffected.
We use the Broadhurst et al. model
as simply an example of a rather severe merging scenario.
This model assumes the number density of the lenses to be
\begin{equation}
   n(\chi_l) = f(\delta t) n_0,
\end{equation}
where $\delta t$ is the lookback time
and the velocity dispersion of the SIS lenses at $\chi_l$ is
\begin{equation}
   \sigma(\chi_l) = \left[ f(\delta t) \right]^{-\nu} \sigma_0,
\end{equation}
where the parameter $\nu$ specifies 
the relation between the mass
of the lenses and their velocity dispersions.
This form implies that if we had $f$ galaxies at lookback time
$\delta t$ each with velocity dispersion $\sigma$ they would
by today have merged into 1 galaxy with velocity dispersion
$[f(\delta t)]^\nu \sigma$. The strength and time
dependence (or redshift dependence) of merging is described by the
function $f(\delta t)$,
\begin{equation}
   f(\delta t) = exp(Q H_0 \delta t),
\end{equation}
where $H_0$ is the Hubble constant and $Q$ represents the merging
rate. The look-back time $\delta t$ is related to $\chi$ through
\begin{equation}
   H_0 \delta t = \Delta^{-1} \int_0^\chi \frac{d\chi}{1+z}.
\end{equation}
We take $Q=4$ (following Broadhurst et al. 1992) and 
$\nu=1/4$ (see Rix et al. [1994] for the
discussion on the value of $\nu$). This choice of parameters
preserves the total probability of lensing and means that
at galaxies $z = 2$ were more numerous by a factor $\sim e^2$
and that their velocity dispersion was smaller by $\sim e^{-1/2}$ than 
those at present with $\Omega=1$.

Since this description of merging depends directly on time 
rather than the redshift, the
function $f$ depends on the individual values of $\Omega$ and
$\Omega_\Lambda$ even in a flat universe. We take $\Omega = 1$ and
$\Omega_\Lambda=0$ as our exemplary flat universe.
The mean separation as a function of source redshift is shown
for this model in Figure 2c as a dotted line. 
The Spearman test shows that the Broadhurst et al. merger model produces a
probability of $P=0.030$, proving that even this strong merging
cannot explain the observed correlation. Most combinations
of $\Omega$ and $\Omega_\Lambda$ in a flat universe have a steep
dependence of $<\Delta\theta>$ at small $z_s$ only, and the
normalized $\Delta\theta$ ranks of the observed
are not significantly affected as long as $\Omega_\Lambda<0.7$. However,
the probability is $P=0.051$ in 
an $\Omega=0.1$, $\Omega_\Lambda=0.9$ universe.
Only the combination of severe merging and extremely large $\Lambda$
(one which would cause severe difficulties with the total
number of lenses as discussed earlier) 
marginally pushes the correlation below the 95\% level.

We also try a less extreme merger model in which the total mass of the
galaxies within a given comoving volume is conserved but the 
comoving number density of galaxies goes like $t^{-2/3}$ while mass 
of individual increases like $t^{2/3}$ where $t$
is the cosmic time since the big bang.
(This is what would be expected
for cosmological infall [Gunn \& Gott 1972] if galaxies grew by
swallowing companion galaxies in an $\Omega=1$ model. It would
overestimate the mass increase in flat and open models with
$\Omega<1$.) We further assume the
mass-velocity relation $M \propto \sigma^4$. This description 
also does not change the total lensing 
optical depth as a function of redshift.
So
\begin{equation}
   n(\chi_l) = n_0 \left[1-(\delta t/t_0)\right]^{-2/3},
   \quad
   \sigma (\chi_l) = \sigma_0 \left[1-(\delta t/t_0)\right]^{1/6},
\end{equation}
where $t_0$ is the current age of the universe.
Again various combinations of $\Omega$ and $\Omega_\Lambda$ 
are tested for a flat universe.
The mean separation for $\Omega=1$, $\Omega_\Lambda=0$ universe 
is shown in Figure 2c as a short-dashed line.
This prescription of merging in any flat universe
produces a probability $P<0.025$ in the Spearman test.

The third model we try is a mass accretion model 
in which the comoving density of
the galaxies is constant but the mass increases with $t^{2/3}$ as in
the cosmological infall model (as would occur if galaxies 
accreted gas by cosmological infall in an $\Omega=1$ model). 
The total mass in galaxies thus increases with time and
the total lensing optical depth is increased:
\begin{equation}
   n(\chi_l) = n_0 (\hbox{constant}),
   \quad
   \sigma (\chi_l) = \sigma_0 \left[1-(\delta t/t_0)\right]^{1/6}.
\end{equation}
Although different combinations of $\Omega$ and $\Omega_\Lambda$
in flat universe give different $\Delta\theta(z_s)$, the difference
is practically negligible (Fig. 2c, dot-dashed line). 
However, the open model produces a 
different $\Delta\theta(z_s)$ (Fig. 2c, long-dashed line)
because the effect due to merging is increased 
to by that due to the curvature. 
The Spearman test for the flat universe
has a probability of $P=0.019$ while that for $\Omega=0.4$, 
$\Omega_\Lambda=0$ open universe $P=0.025$. 
So even moderate mass accretion in an open universe can not produce
the strong correlation seen in the data.

\section{Clusters}

Since large image separations in some lens systems 
($\Delta\gtrsim5\arcsec$) are too large
to be explained comfortably within the currently accepted galaxy
mass distributions, we expect these systems to be the
result of galaxy lensing aided by a cluster as in the case of 0957. 
We investigate what kind of
effects would be expected if lensing is aided by a cluster.
The cluster is simply modeled as a sheet constant mass surface
density (TOG). 

When multiple images are produced by an SIS lens aided by a cluster,
the lensing cross section is not affected but the image separation
is widened (TOG),
\begin{equation}
   \frac{\Delta\theta_{G+C}}{\Delta\theta_G} 
   = \left( 1 - \frac{\Sigma}{\Sigma_{cr}} \right)^{-1},
\end{equation}
where $\Delta\theta_{G+C}$ is the separation by the SIS plus cluster
and $\Delta\theta_G$ that by the SIS alone,
$\Sigma$ is the surface mass density of the cluster, and
$ \Sigma_{cr} \equiv \Sigma_0 S(\chi_s)/[S(\chi_l)S(\chi_s-\chi_l)] $
is the critical surface mass density, 
where $\Sigma_0 \equiv c^2/(4\pi G a_0)$.
Thus the total lensing probability is unchanged, 
but the mean image separation is
\begin{equation}
   <\Delta\theta> = 2 \alpha_0
   \frac
   {\int_{0}^{\chi_s} 
   \frac{\Sigma_{cr}}{\Sigma_{cr}-\Sigma}
   \frac{S^3(\chi_s-\chi_l)S^2(\chi_l)}{S^3(\chi_s)} d\chi_l} 
   {\int_{0}^{\chi_s} 
   \frac{S^2(\chi_s-\chi_l)S^2(\chi_l)}{S^2(\chi_s)} d\chi_l}.
\end{equation}

If one attributes the large separation lenses seen at small
source redshift to a cluster helping a galaxy one might hope that
the observed effect is due to a lack of clusters at large
redshifts. Can this be due to an evolution of clusters with 
redshift? No. Because nearby clusters help lensing for all
more distant sources and even more effectively as the source
redshift increases. If there were no distant clusters beyond
some redshift $z_i$ then this would have the effect of causing
an {\em increase} in image separation with increasing source
redshift. 

We assume two cases for the position of the cluster. 
For first, we assume the same redshift 
for the cluster and the lensing galaxy. 
The resulting mean image separation is shown in Figure 2d for
an $\Omega=1$, $\Omega_\Lambda=0$ universe (dotted line). The mean separation
increases with source redshift because adding the cluster
effectively makes the mass distribution even more extended
than SIS. For the second, the redshift
of a cluster is at some fixed value smaller than the source 
redshift. The mean image separation in the same flat universe
for this case is shown in Figure 2d for the cluster redshift
of 0.5 (dashed line). 
It is also an increasing function of source redshift.
This is expected because $\Sigma_{cr}$ for any cluster
is always smaller
for a higher redshift source regardless of the lens redshift.
Therefore, for a given surface density, a cluster is closer
to the critical surface density for more distant sources, and we expect
larger image separations. This is just the opposite of the
correlation seen in the data. 
   
\section{Other Implications}

\subsection{Test of the Curvature of the Universe}

It was hoped that the dependence of image separations of lens systems
on the redshift of the source may make it possible to
test the curvature of the universe directly (TOG; GPL).
However, the small number of the lens systems available
makes this test very difficult (GPL).
Here we examine how many multiple image lens systems
are required to reliably test the curvature of the universe.
Since we are not sure that the observed distribution of the image 
separations, especially that of the large separation ones,
is explained by a simple lensing model
where the sources are lensed by a single galaxy 
following the Schechter luminosity function,
we do not use any assumptions on the lensing galaxies and use
only the observed image separation distribution as the
intrinsic distribution we are likely to discover in the future.

Although the observed data may contain the curvature effect
already, we assume that the observed distribution is just the intrinsic
one before being affected by the curvature. We create
$N$ Monte-Carlo multiple images systems out of randomly
shuffled images separations and source redshifts seen 
in the observed lens systems (and listed in Table 1).
This shuffled data set will have the same histogram of separations
as observed and the same histogram of observed redshifts---but
the redshifts and separations will be by definition
uncorrelated as would be true in a flat model with SIS lenses.
Then the image separations of the simulated samples
are multiplied by the mean image separation at the simulated
redshift expected in various cosmologies.
We then run the Spearman test on all simulated data sets
to detect the existence of the negative correlation
at above the 95\% confidence level. We find that to distinguish 
the flat universe versus the empty universe at the 95\% probability level
requires 800 multiple image systems. Proving a less extreme open universe
like the $\Omega=0.4$, $\Omega_\Lambda=0$ universe at the same
95\% confidence level requires a staggering
$\sim 1600$ systems. This proves that pure curvature test from lens
statistics is harder than originally expected mainly because
the observed scatter in image separations
is larger than initially expected. Yet it might well be
within the reach of future sky surveys (Sloan Digital Sky Survey
expects to discover $\sim100$ new lenses in its spectroscopic
survey, and about 1000 new lenses from its faint quasar candidate list
based on their stellar type images but QSO type colors.
This is how many such lenses would be expected to be confirmed by
later spectra from these candidates using other telescopes.
[SDSS Collaboration: NASA Proposal 1997]).

\subsection{New Limits on $q_0$}

GPL discovered that in a $\Lambda \neq 0$ universe 
where the observer's antipode is within the particle horizon, a source just
beyond the antipode is over-focused due to the lensing action of the universe
as a whole and cannot create multiple images under most lensing
mass distributions, e.g., SIS, SIS with external shear, 
and elliptical potential.
Hence, the existence of ordinary multiple image lens
systems at various source redshifts up to some maximum in general
constrains the antipode to be farther away than the
largest observed redshift multiple lens system source
(now at $z_s=4.5$). (See GPL for details.) 
This limit on the antipodal redshift
(now $z_{antipode} > 4.5$) revises the allowed 
parameter space in $\Omega$ vs. $q_0$ (the unshaded region in Fig. 3).
We also provide a graph (Fig. 4) for the lower limit on
$q_0$ as a function of the antipodal redshift, so as new
record breaking (in $z_s$) lensed QSO's are discovered, the lower
limit on $q_0$ can be revised upward accordingly. 

\section{Summary and Discussion}

We find that the currently observed multiple image lens systems 
show a very strong negative correlation 
between the image separation and 
the redshift of the source in the sense that larger redshift sources
have smaller separations. The probability of this occurring
in a flat universe with standard non-evolving galaxies is only 1\%.

Possible causes are investigated: the curvature of the universe, 
different mass profiles for lensing galaxies, merger or accretion of 
galaxies, and lensing aided by clusters. Although all of these except
the lensing-aided-by-clusters model can create a negative correlation
between the separations and the source redshifts,
none of them produce a negative correlation as strong as
that seen in the data. This leaves us with a puzzle.
If there is a cause (not explored in this work) that can explain the
correlation, it has to
have very strong evolutionary effects, especially between $z \sim 2$
and $z \sim 4$. 

Interstellar dust in younger galaxies causing obscuration is not
helpful. Obscuration might prevent us from seeing some high 
redshift QSO's (see Malhotra, Rhoads, \& Turner 1996 about 
the evidence for dusty gravitational lenses). 
But in analyzing the separation versus redshift
question we are only dealing with the ones we do see. If there
is dust in the lensing galaxies we would expect it to knock out
small separation cases preferentially 
and if dust increases with increasing 
redshift in lensing galaxies as we would expect, then this
would cause separations to increase slightly with increasing source
redshift, which is the opposite of what we observe.

Are there any observational selection effects that would produce
the effect? It is not easy to think of one. One of the small
separation large redshift cases (1208) was discovered with the
Hubble Space Telescope which is better able to 
discover small separation cases than ground based telescopes.
But of course the Space Telescope is equally well able to
discover small source redshifts. The HST snapshot survey
includes both small and large separation cases and
both small and large redshift cases: 0957, 0142, 1115, 1413,
1208, and 1120 (Maoz et al. 1993). 
As a matter of fact, even this small number
of systems shows a very strong correlation 
(a two-sided probability that they are drawn 
from a random data sample in a flat universe is $P=0.036$). 
Many lenses are found by the VLA where
the source redshift is found only after the confirming
spectra are taken. The VLA can detect separations as small as
$0\farcs3$ and QSO's at any redshift. 
Optical surveys simply stumble on cases and might
miss some small separation cases but again would be expected
to find large and small source redshifts equally well. 
Therefore, it is hard to think of a selection effect that
would be biased against detection of large separation, large redshift
quasars only and which would be present in HST observations,
VLA observations, and ground based optical observations. 

One possible, yet unlikely, explanation of the 
observed correlation may be ``false'' gravitational lenses.
We have shown that if 
three or more of the largest separation cases 
or two or more of the largest redshift (small separation) cases turn
out not to be true lenses, the probability of having as strong
a correlation as that seen in the data 
increases above 5\% level in a flat universe. Smaller number of
``false'' cases would do the same for an open universe. 
However, most of the lens cases in Table 1
have been on the list for more than a few years without being
disconfirmed. Also, even if the probability is above 5\%,
it has to be multiplied to the
additional likelihood that the specific cases are ``false'' lenses
and the final probability would be very small.
So it seems unlikely that three or more
large separation cases or two or more large redshift cases will
turn out to be just physical pairs and at the same time 
the correlation is from random distribution. However, the
likelihood would be larger in an open universe: A single ``false''
lens (1208) and $\sim 6\%$ of chance in an $\Omega=0.4$, 
$\Omega_\Lambda=0.0$ universe could produce a correlation
as strong as that seen in the data.
Needless to say, either stronger confirmation
or disconfirmation of large separation cases and large redshift cases 
through future observations would be very helpful.

If this negative correlation is real (i.e., not just a statistical fluke,
or an observational selection effect, 
or due to``false'' lens contamination)
we may have to revise various conclusions.
For example the limit on $\Omega_\Lambda$
(Fukugita \& Turner 1991; Fukugita et al. 1992; Kochanek 1996)
may have to be weakened because its hypothesis (constant comoving density
unevolving SIS lenses, flat universe) would be wrong. On the other hand,
if the current sample is a statistical fluke, then the observed lens
systems must constitute a non-typical subset of the parent
population and the limit of $\Omega_\Lambda$ would also have to be
weakened for the reason that our current sample is not a fair sample.
The same thing can be said if the correlation is due to
unknown observational selection effects or if the lens sample
is contaminated with ``false'' cases.
It is worth noting that if the negative correlation is just
a statistical fluke we would
be twice as likely to observe such an anomaly in an open
universe with $k=-1$ and $0\leq\Omega\leq0.4$ as we would be
to observe it in a flat $k=0$ universe with 
$\Omega+\Omega_\Lambda=1$.

Finally, the existence of multiple image systems up to a redshift 
of 4.5 places a limit on the deceleration parameter, $q_o > -2.0$.

\acknowledgments

We thank E. Turner for a very useful discussion.
This work was in part supported by Korea Science Foundation
grant 95-1400-04-01-3, NSF grant AST9529120, AST9424416 
and NASA grant NAG5-2759.
MGP's visit to Princeton University during which most of the
work was done was supported by the Korea Research Foundation.


\newpage

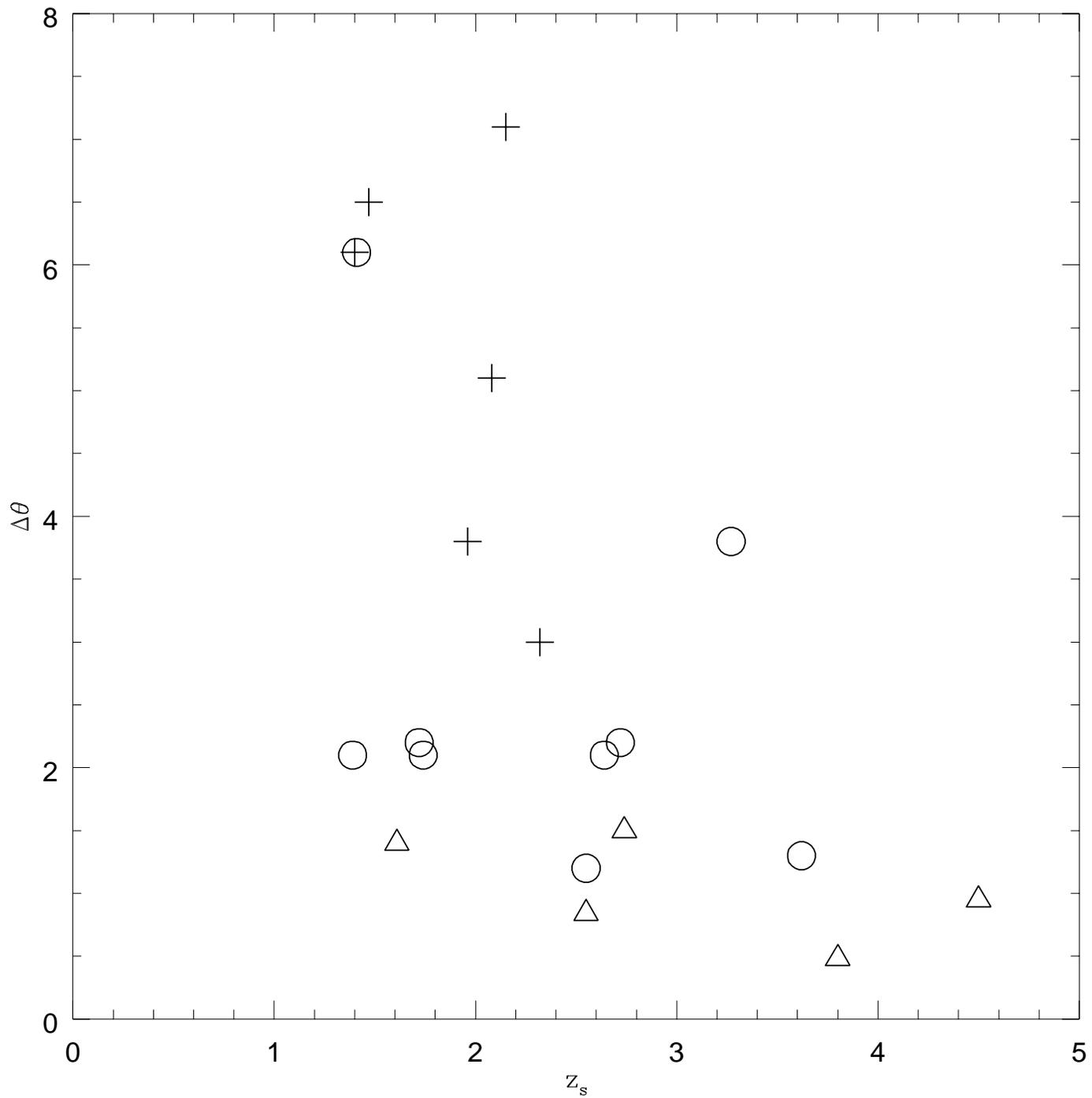
\begin{figure}
  \caption{Maximum image separation $\Delta\theta$ vs. 
           source redshift $z$ of 
           multiple image gravitational lens systems.
           Circles denote class A
           (``I'd bet my life this is a lens.''), triangles
           class B (``I'd bet your life this is a lens.''), 
           and crosses class C
           (``You should worry if I'm betting your life.''),
           according to Keeton \& Kochanek (1996).}
\end{figure}

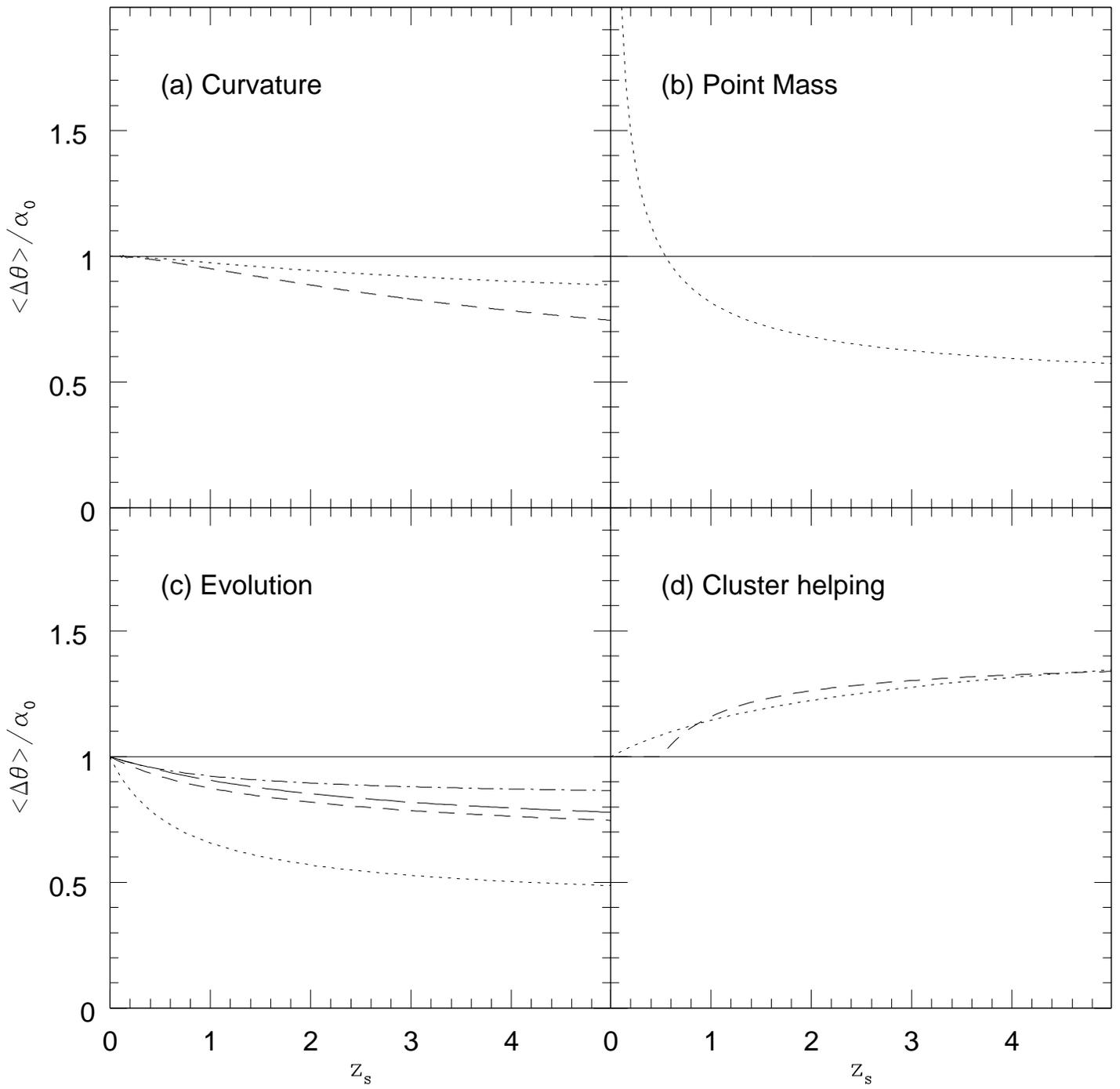
\begin{figure}
  \caption{Mean separation of images as a function of source
           redshift for various possiblities: (a) Curvature:
           Flat universe ($\Omega+\Omega_\Lambda=1$)
           (solid line), $\Omega=0.4$ open universe (dotted),
           and empty universe, $\Omega=0$ (dashed), all with SIS
           lenses. (b) SIS lenses (solid)
           vs. point mass lenses (dotted, in arbitrary
           unit) in a $\Omega=1$ flat universe.
           (c) Evolution: No mergers (solid), 
           Broadhurst et al. merger model (dotted), cosmological
           infall of satellite galaxies (short-dashed), 
           mass accretion (dot-dashed), all in a flat universe,
           and mass accretion (long-dashed) in an $\Omega=0.4$ 
           open universe. (d) Lensing aided by a cluster
           at the same redshift as the galaxy (dotted line) 
           and at the fixed redshift of 0.5 (dashed) in a
           flat universe.}
\end{figure}

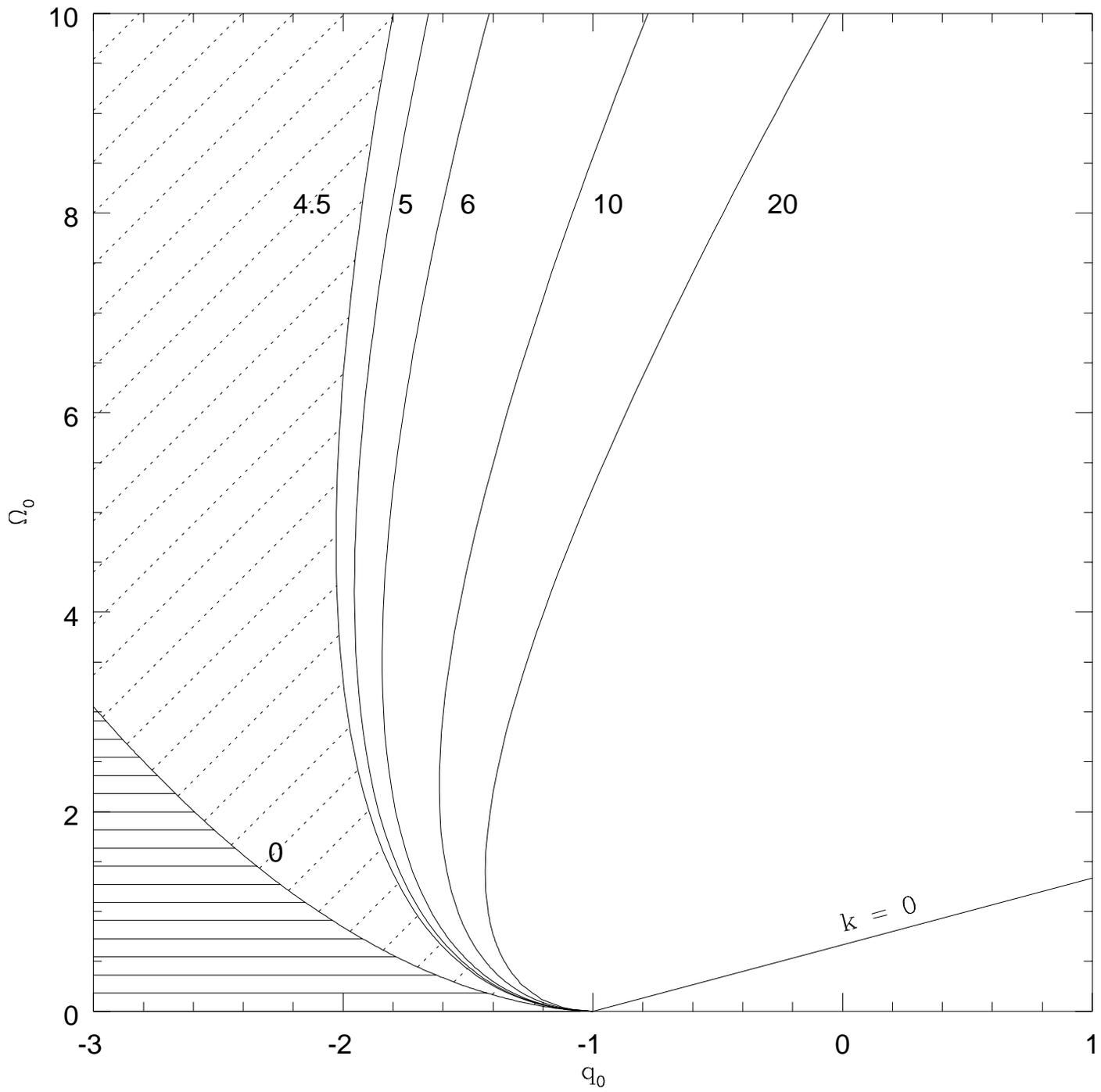
\begin{figure}
  \caption{The lines of constant antipodal redshift in $\Omega$
           vs. $q_0$ plane. The numbers
           denote the value of the antipodal redshift $z_{antipode}$.
           There is no big bang in the horizontally shaded region 
           below $z_{antipode}=0$. The horizontally and diagonally shaded
           regions are both excluded if $z_{antipode}>4.5$ as must be the
           case since we see numerous multiply lensed QSO's up
           to and including one at $z=4.5$ (see GPL). The solid
           line marked $k=0$ represents flat universes, 
           $\Omega+\Omega_\Lambda=1$.}
\end{figure}

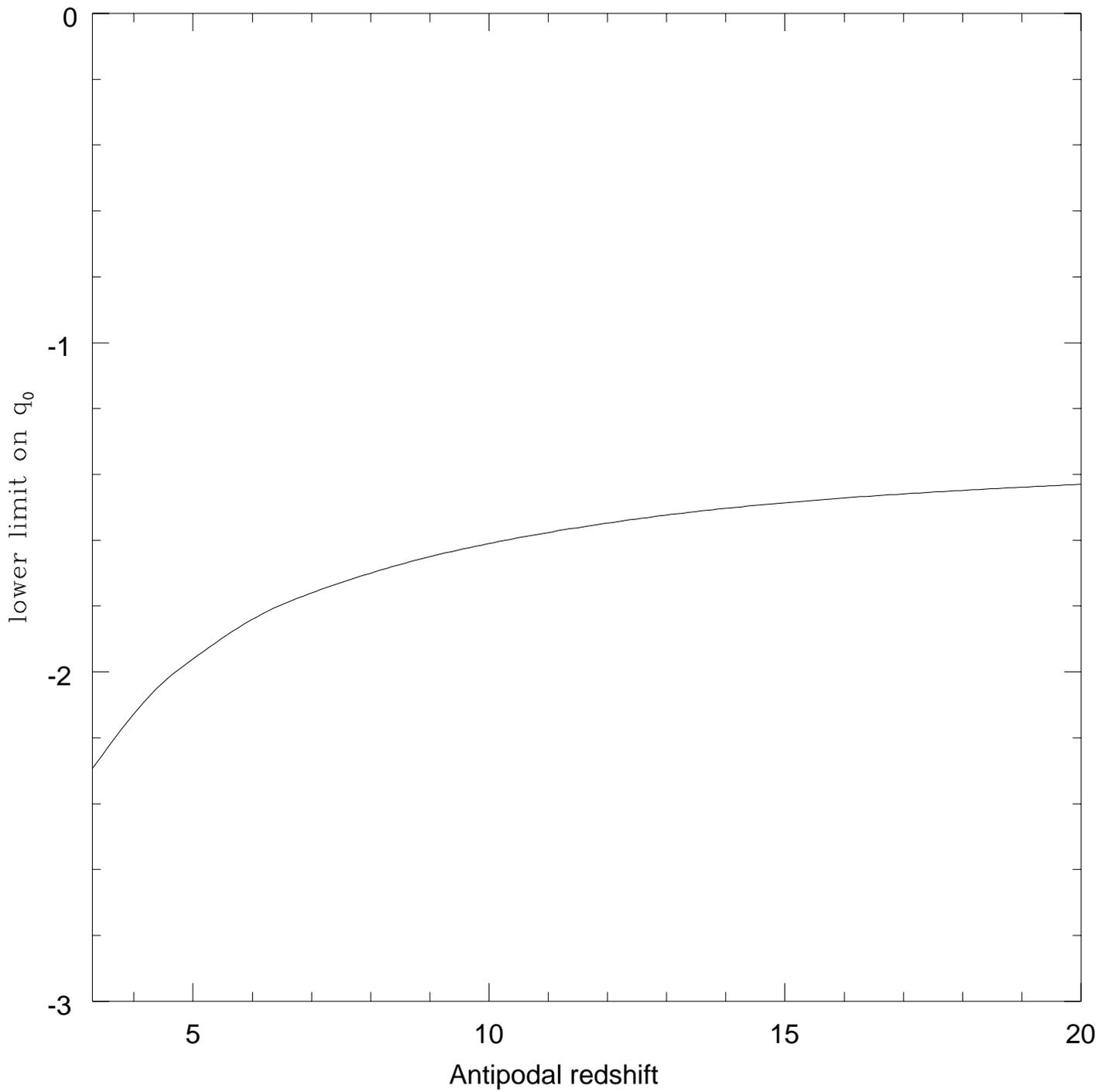
\begin{figure}
  \caption{Lower limit on $q_0$ as a function of antipodal redshift.}
\end{figure}

\clearpage
\begin{table*}
\begin{center}
\begin{tabular}{llll}
Name & {$z_s$} & {$\Delta\theta$} 
     & {References} \\
\tableline
CLASS1608+656	& 1.39	& 2.1	& Myers et al. 1995 \\
QJ0240-343	& 1.4	& 6.1	& Tinney 1995 \\
0957+561	& 1.41	& 6.1	& Walsh et al. 1979 \\
1120+019	& 1.47	& 6.5	& Meylan \& Djorgovski 1989 \\
CLASS1600+434	& 1.61	& 1.4	& Jackson et al. 1995 \\
1115+080	& 1.72	& 2.2	& Weymann et al. 1980 \\
MG1654+1346	& 1.74	& 2.1	& Langston et al. 1989 \\
1634+267	& 1.96	& 3.8	& Djorgovski \& Spinrad 1984 \\
1429-008	& 2.08	& 5.1	& Hewett et al. 1989 \\
2345+007	& 2.15	& 7.1	& Weedman et al. 1982 \\
HE1104-1805	& 2.32	& 3.0	& Wisotzki et al. 1993 \\
J03.13		& 2.55	& 0.84	& Claeskens et al. 1996 \\
H1413+117	& 2.55	& 1.2	& Magain et al. 1988 \\
MG0414+0534	& 2.64	& 2.1	& Hewitt et al. 1992 \\
0142-100	& 2.72	& 2.2	& Surdej et al. 1987 \\
LBQS1009-0252	& 2.74	& 1.5	& Surdej et al. 1994 \\
2016+112	& 3.27	& 3.8	& Lawrence et al. 1984 \\
B1422+231	& 3.62	& 1.3	& Patnaik et al. 1992 \\
1208+1011	& 3.80	& 0.48	& Bahcall et al. 1992; \\
                &       &       & Magain et al. 1992 \\
BRI0952-0115	& 4.5	& 0.95	& McMahon et al. 1992 \\
\end{tabular}
\end{center}

\tablecomments{
(1) $z_s$: Redshift of the source. 
(2) $\Delta\theta$: The maximum image separation.
(3) See Surdej \& Soucail [1994] or
    Keeton \& Kochanek [1996] for more references.}
\caption{Multiple image lens systems used in this work.}
\end{table*}

\clearpage
\begin{table*}
\begin{center}
\begin{tabular}{ll}
Models & Probability \\
\tableline
Flat universe ($\Omega + \Omega_\Lambda = 1$) 	& .012 \\
Empty universe ($\Omega=0$, $\Omega_\Lambda = 0$) 	& .030 \\
Open universe ($\Omega=0.4$, $\Omega_\Lambda = 0$)	& .019 \\
Point mass lens in a flat universe 			& .030 \\
Merger model in a flat universe\tablenotemark{b}
                       					& $<.05$ \\
Cosmological infall in a flat universe universe 
                       					& $<.03$ \\
Mass accretion in a flat universe\tablenotemark{c} 
                       					& .019 \\
Mass accretion in an open universe\tablenotemark{d}
                       					& .025 \\
Lensing aided by a cluster				& $<.01$ \\
\end{tabular}
\end{center}
\tablenotetext{b}{Broadhurst et al. merger model in a flat 
                  universe with $\Omega_\Lambda<0.9$.}
\tablenotetext{c}{Flat universe with $\Omega=1$, $\Omega_\Lambda=0$.}
\tablenotetext{d}{Open universe with $\Omega=0.4$, $\Omega_\Lambda=0$.}
\caption{Two-sided probabilities of observing a correlation
         as strong as that seen in the data in various models, 
         using the Spearman rank correlation test}
\end{table*}

\end{document}